\documentclass[floatfix, prb, aps,amsmath,amssymb,  twocolumn,superscriptaddress,10pt,showkeys, author-year]{revtex4-2}

\usepackage[usenames,dvipsnames]{xcolor}
\usepackage{hyperref}
\hypersetup{
colorlinks=true,
citecolor=NavyBlue,
linkcolor=NavyBlue,
filecolor=NavyBlue,
urlcolor=NavyBlue
}
\usepackage{graphicx}
\usepackage{dcolumn}
\usepackage{bm}
\usepackage{amsmath, amsfonts, amssymb}
\usepackage{orcidlink}

\usepackage{comment}

\usepackage[normalem]{ulem}

\begin{document}
\title{Self-assembled clusters of magnetically tilted dipoles}

\author{P. D. S. de Lima\orcidlink{0000-0002-7353-536X}}
\affiliation{School of Physics, Trinity College Dublin, Dublin 2, Ireland}%
\affiliation{Departamento de Física Teórica e Experimental, \\Universidade Federal do Rio Grande do Norte, 59078-970 Natal-RN, Brazil}
\author{A. Lyons}
\affiliation{School of Physics, Trinity College Dublin, Dublin 2, Ireland}%
\author{A. Irannezhad}
\affiliation{School of Physics, Trinity College Dublin, Dublin 2, Ireland}%
\author{J. M. de Araújo\orcidlink{0000-0001-8462-4280}}
\affiliation{Departamento de Física Teórica e Experimental, \\Universidade Federal do Rio Grande do Norte, 59078-970 Natal-RN, Brazil}
\author{S. Hutzler\orcidlink{0000-0003-0743-1252}}
\affiliation{School of Physics, Trinity College Dublin, Dublin 2, Ireland}%
\author{M. S. Ferreira\orcidlink{0000-0002-0856-9811}}
\affiliation{School of Physics, Trinity College Dublin, Dublin 2, Ireland}%
\affiliation{Centre for Research on Adaptive Nanostructures and Nanodevices (CRANN) \& Advanced Materials and Bioengineering Research (AMBER) Centre, Trinity College Dublin, Dublin 2, Ireland}

\date{\today}

\begin{abstract}
Motivated by the idea of using simple macroscopic examples to illustrate the physics of complex systems, we modify a historic experimental setup in which interacting floating magnets spontaneously self-assemble into ordered clusters. By making the cluster components mechanically stable against their natural tendency to flip and coalesce, we can monitor the torque experienced by individual magnets through the macroscopic tilt angles they acquire when exposed to non-collinear external fields. A mathematical model that reproduces the empirical observations is introduced, enabling us to go beyond the experimental cases considered. The model confirms the existence of alternative orderings as the number of objects increases. Furthermore, a simpler and more mathematically transparent version of our model enables us to establish the conditions under which the cluster structure is stable and when it will collapse. We argue that our simple experimental setup combined with the accompanying models may be useful to describe general features seen in systems composed of mutually repulsive particles in the presence of modulating confining fields.
\end{abstract}

\maketitle


\section{Introduction}

It is well known that aggregates of interacting particles may order spontaneously when placed under the influence of confining forces. Equal-volume hard spheres, which are normally disordered in bulk, crystallise into ordered structures when placed inside cylinders of a diameter slightly exceeding the sphere diameter~\cite{MughalEtal2012, WinkelmannChan2023}. A similar behaviour is found for (soft) bubbles~\cite{MeagherEtal2015,TobinEtal2011,WinkelmannChan2023}. Likewise, the case of ions in the presence of a confining electric field~\cite{LaiLin1999,NazmitdinovEtal2017, MughalEtal2023,MughalEtal2023calligraphy} is yet another example of mutually repulsive particles spontaneously self-assembling into ordered structures when spatially constrained. 
The list of examples with similar behaviour is quite extensive, appearing over a wide range of length scales~\cite{self_assembly_multiscale}.

One of the earliest attempts of 
elucidating the effect of confinement-induced ordering in clusters dates back to Alfred Marshall Mayer who in 1878 ``devised a set of macroscopic experiments which illustrates the action of atomic forces, and their arrangements in molecules’'~\cite{Mayer1878a}. Mayer magnetised sewing needles with their tips having the same polarity and mounted them on small corks, which were placed in a bowl filled with water. When the opposing pole of a large magnet is placed above the mutually repelling needles, they float towards each other and quickly arrange themselves in some regular pattern. Mayer identified different configurations using up to 51 needles which arranged themselves in rings~\cite{Mayer1878b}; this included various {\em alternative} configurations found for the same number of needles and was subsequently instrumental in providing insightful analogies for the shell structure formations at the atomic level~\cite{Bragg1925,Snelders1976}.



Clusters of magnetic objects have been the focus of scientific attention for decades~\cite{PhysRevE.65.061405, PhysRevLett.85.5464, MARTINEZPEDRERO,doi:10.1021/acsnano.0c09952} but interest has increased more recently due to applications in environmental science~\cite{ocean-cleaning}, 
medicine~\cite{Ulbrich2016, Nguyen2021, pharmaceutics15071872, Ulbrich2016, Nguyen2021, pharmaceutics15071872}, magnetic memory devices~\cite{nano10071318} and quantum computing~\cite{9076325, Liao2023}, to name but a few. Obviously, this calls for an understanding of magnetic clusters that goes beyond the simple formation of ordered structures. For example, determining the cluster magnetic texture, {\it i.e.}, how individual magnetic moments are dispersed both in magnitude and direction, can help explain how these systems respond to external fields. Therefore, despite being over a century old, Mayer’s seminal idea of searching for a macroscopic analogue able to capture features that may be prevalent across different length scales remains relevant.


Numerous experiments of that kind have been proposed over the years~\cite{mag_cluster_ex1, mag_cluster_ex2} but the information contained in the magnetic angle alignment in clusters has often been discarded or ignored. Mayer himself did not comment on any observed deviations of his magnetised needles from the vertical direction. Nemoianu {\em et al.}~\cite{NemoianuEtal2022} attribute this to the buoyancy force provided by the cork stoppers to which the needles were attached. In their own recreation of Mayer's experiment, small ``pill-like'' cylinder magnets are glued to flat-bottom plastic cups; again this will hinder any magnetic tilt. Also in the experimental study by Ref. \cite{RiverosEtal2004} tilt is deliberately suppressed;  the magnets were stuck underneath floating cork disks to have the centre of gravity below that of buoyancy. To be consistent with their own (and Mayer's) experimental set-up, and to avoid a further complication of their equations of force equilibrium, Nemoianu {\em et al.}~\cite{NemoianuEtal2022} also restricted their numerical study of magnet configurations to the case of vertically aligned magnetic moments.

The magnetic tilt of individual objects is a degree of freedom that adds richness to the problem of self-assembled magnetic clusters and should not be discarded or ignored. With that in mind, here we adapt the floating-magnets experiment so that the magnetic moment of cluster components is permitted to tilt partially instead of being entirely suppressed by buoyancy. By attaching additional weight to the bottom of floating spheres, we make them mechanically stable against the natural tendency that dipole moments have of flipping and attaching to one another. In this way we preserve the repulsive force between the floaters; at the same time we allow it to be modulated by a non-collinear confining field (see Sec. \ref{s:experiments} for details). Furthermore, we introduce a theoretical model that describes the details of this modified experimental setup. The model displays excellent agreement with the experimental measurements, which allows us to go beyond our observations and generate a phase diagram describing how the number of rings within a cluster depends on the total number of objects, as well as on the tilting angle of an externally applied magnetic field. Our findings confirm the existence of alternative orderings as the number of objects increases. Furthermore, the model can be used to establish the conditions under which the cluster structure is stable.

The sequence adopted in this manuscript is as follows. In the next section, we present the modified experimental setup that was used to generate the ordered structures of the magnetic cluster, including the details of how the positions of the magnetic objects and their respective tilting angles were measured. This is followed in Sec. \ref{model} by the introduction of a mathematical model that describes through an energy balance equation how the system self-assembles into rings. By making use of a few simplifying approximations introduced in Sec. \ref{esrm} we are subsequently able to obtain expressions for the ring radii, infer about its magnetic texture and address the robustness of the cluster structure against symmetry-breaking fluctuations.

\section{Experiment}
\label{s:experiments}

\begin{figure*}[!htpb]
    \centering
   (a)\includegraphics[width=0.6\linewidth]{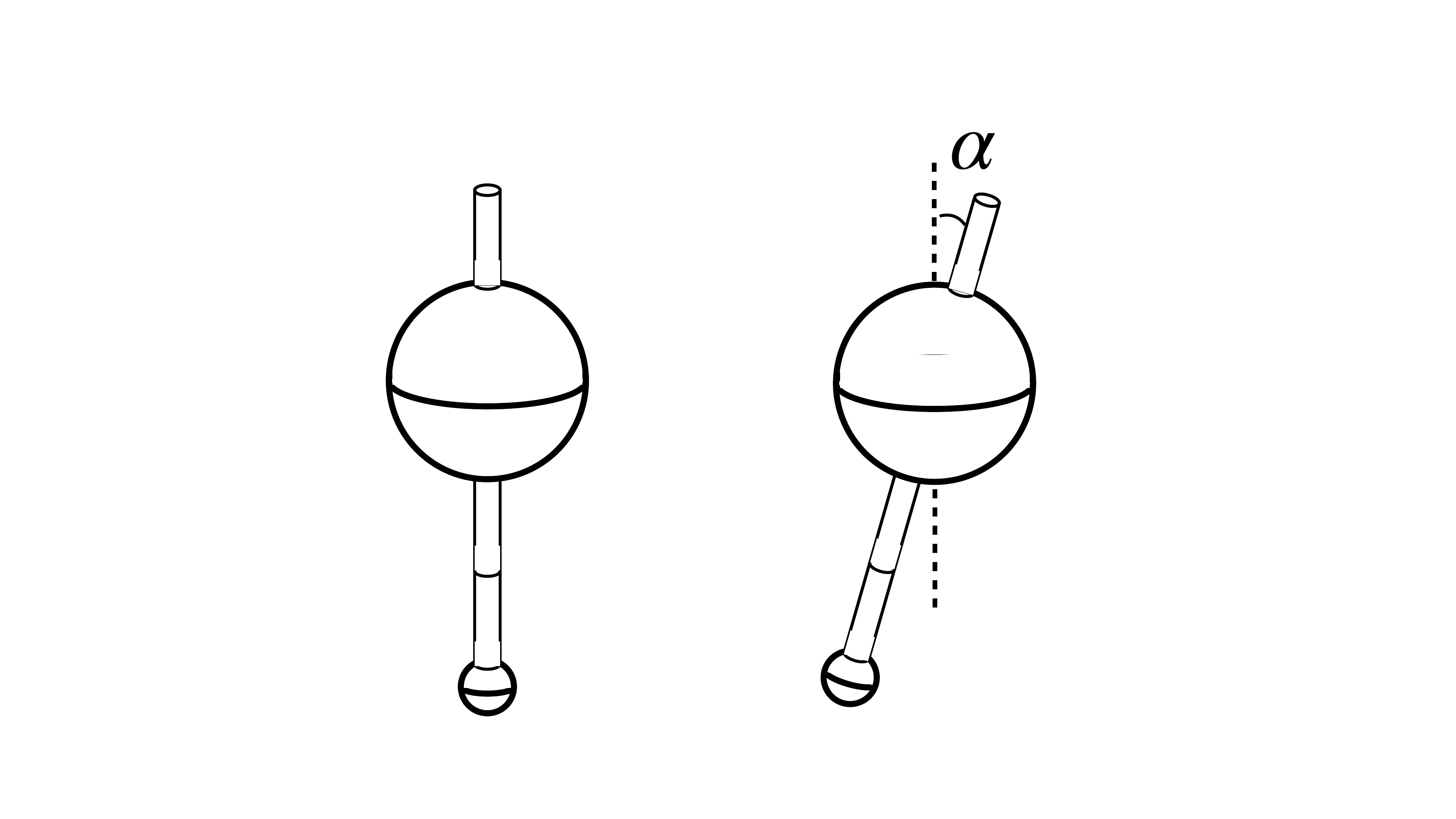}
   (b)\includegraphics[width=0.6\linewidth]{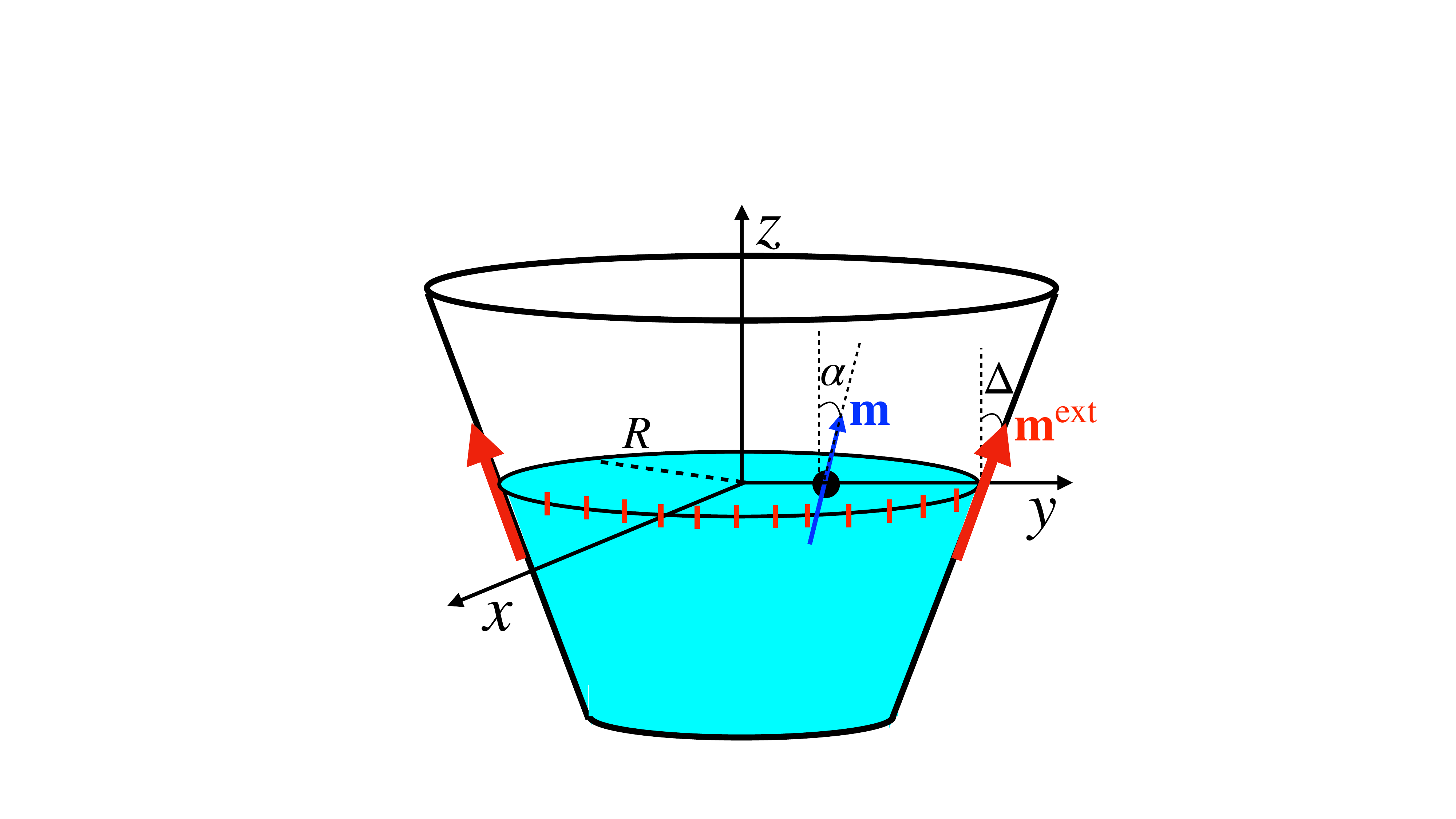}
   (c)\includegraphics[width=0.6\linewidth]{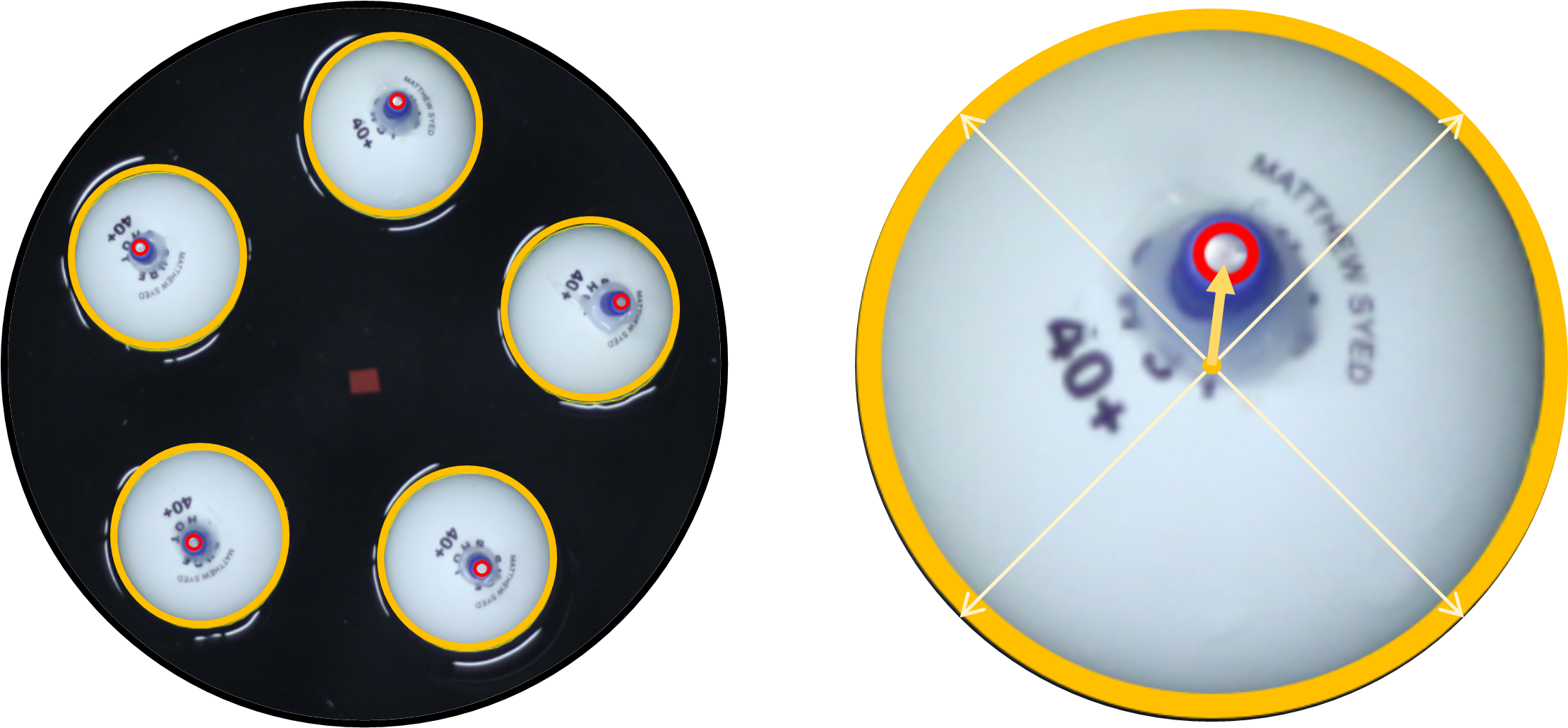}
\caption{Experimental set-up and imaging. (a) Schematic drawing of the floating object which is made up of a ping-pong ball (diameter of 4 cm) onto which bar magnets of length 2.7 cm are glued: one at the top and two at the bottom. The curved lines around the circumference represent the water level. The metal sphere attached at the end of the bottom magnet serves to increase its stability against flipping over when placed in water. Under the influence of a magnetic torque, these ``fishing floats'' tilt by an angle $\alpha$. (b) Schematic diagram of the bucket. The slanted side wall of the bucket forms an angle $\Delta=17^{\circ}$ with the vertical direction. The radius is $R = 15$ cm at the air-water interface where the floats are located in equilibrium. A black dot indicates one floating sphere with its magnetic dipole moment $\mathbf{m}$ represented by a blue vector that is radially tilted by $\alpha$. The red arrows $\mathbf{m}^{\mathrm{ext}}$ on opposite sides of the bucket represent the dipole moments of two of the magnets responsible for generating the confining forces. Equally spaced red tick-marks around the circumference of the water surface indicate the positions of the other magnets attached to the bucket wall. (c) Photographic image of the top view, showing an equilibrium arrangement for five floating spheres. Their individual perimeters are highlighted in yellow. Red circles identify the perimeter of the tip, which is used to determine the angle of inclination, also indicated by the yellow arrow. }
\label{f:set-up}
\end{figure*}

Our macroscopic systems of floating magnets were realised experimentally by glueing two bar magnets on opposite sides of ping-pong balls, as shown in Fig. \ref{f:set-up}(a). To provide stability against flipping over when placed in water, an additional magnet, together with a metal sphere, was added on the water-immersed side. The floating magnets were placed into a water-filled bucket, shaped as the frustum of a cone, onto whose outer perimeter, at the water level, we attached a total of 70 equally spaced magnets, as shown schematically in Fig. \ref{f:set-up}(b). The effect of the confining magnetic field, produced by these outer magnets, is that a single object, placed anywhere in the bucket, will float to take on its equilibrium position in the centre. Systems of several floating magnets, whose repelling forces, in the absence of the confining field, drive them to locations along the perimeter of the bucket, float to equilibrium positions around the centre in the presence of the field (see Fig. \ref{f:set-up}(c)). 

A camera mounted on a tripod was placed above the bucket and vertically aligned with its centre. The positions of the floating spheres and the inclination of the magnets away from the vertical were then determined as follows. We first located the centre of the confining field. This can differ from the geometric centre of the bucket due to inhomogeneities in the spacing of the magnets which produce this field. A single floating magnet was placed into the bucket and allowed to float until it came to a complete stop. The camera was then centred onto the top of the ping-pong ball using crosshairs on the camera's viewfinder and an image was captured. One by one, more magnets were placed onto the water surface; each magnet was left to settle into its respective equilibrium position for about a minute, with any water movement having stopped at this stage. This final position was then captured on camera.

Image analysis was carried out using the open-source software ImageJ \cite{ImageJ2021}. Circles were drawn around the perimeters of the spheres, enabling us to locate the exact coordinates of their respective centres. The same procedure was employed for the circular top of each of the magnets pointing out from the balls; these coordinates were then used to compute the tilting experienced by each one of the floating spheres.  Fig.  \ref{f:set-up}(c) illustrates this procedure.

For some numbers of magnets (exceeding four) multiple arrangements were possible. These were achieved by manually disturbing an equilibrium configuration before placing an additional ball. The maximum number of floating magnets that we examined was 17.  Most of the configurations that we found were those reported by~\cite{Mayer1878b} and/or~\cite{NemoianuEtal2022}. See Appendix \ref{app-structures} and Tab. \ref{tab:structures} for a list of all the structures for up to 17 magnets.

\section{Model and Experimental Validation}
\label{model}

\begin{figure*}[!htbp]
    \centering
    \includegraphics[width=\textwidth]{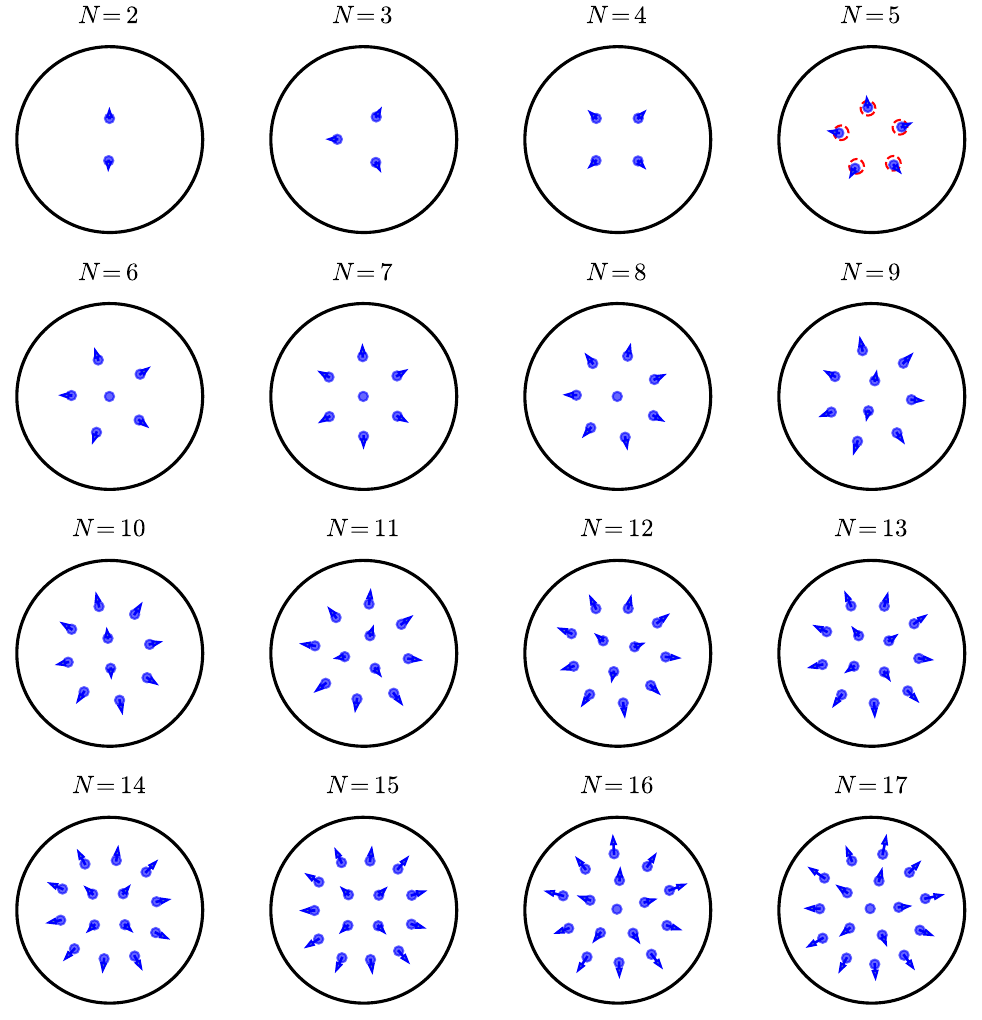}
    \caption{Top view of the equilibrium configurations for $N$ floating objects, from $N=2$ to $N=17$. Solid circles represent the bucket wall, which forms an angle $\Delta=17^\circ$ with the vertical direction. Blue dots mark the central positions of the floaters while the arrows indicate their angles of tilt (arrows are omitted for angles less than $0.5^\circ$). These values are found through the GSA-minimisation of Eq. (\ref{eq:tot_energy_gsa}). The red dashed-line circles drawn in the case $N=5$ indicate the observed positions in Fig. \ref{f:set-up}(c) and illustrate the agreement with the experimental observations.}
    \label{fig:mag_cluster_delta_17}
\end{figure*}

\begin{figure}[!htbp]
    \centering
    \includegraphics[width=3.4in]{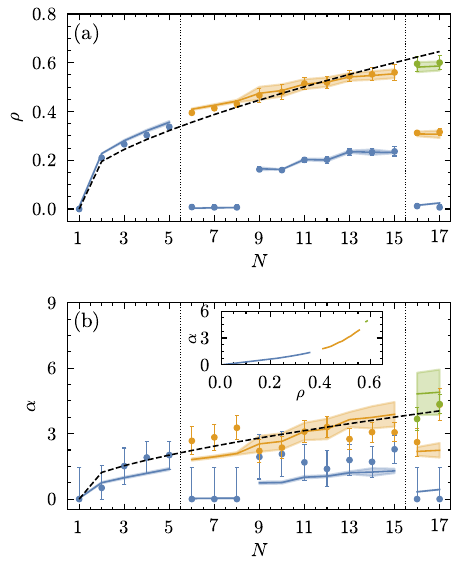}
    \caption{Comparison between the GSA-generated results and the experimental observations for (a) ring radii as a function of the number of floating objects $N$; (b) Tilt angle $\alpha$ (in degrees) as a function of $N$. Two vertical dotted lines separate the regions with different numbers of rings: the leftmost part ($N < 6$) corresponds to cases where only a single ring is found; the middle ($6\leq N <16$) is for two rings and the right ($N \geq 16 $) is for three. The colour code is such that blue refers to the innermost ring, orange to the second ring and green to the third ring. Symbols indicate the experimentally measured values (with error bars) while the simulated results (at integer values of $N$) are connected by solid lines which serve as a guide to the eyes. The shaded regions indicate the standard deviation observed in the simulations. The black dashed lines in both panels correspond to results for the outermost ring using an effective single-ring model (see Sec. \ref{esrm}). The inset depicts the GSA results of $\alpha(\rho)$ for the outermost ring and is a representation of the magnetic texture seen in these clusters.}   \label{fig:exp_validation}
\end{figure}

We start by identifying and labelling the key quantities in the problem. We consider $N$ floating spheres within a container that has a circular cross-section with radius $R$ at the air-liquid interface (see a schematic diagram of the bucket in Fig. \ref{f:set-up}(b)). It is convenient to identify the position $\mathbf{r}_j$ of each sphere in polar coordinates, {\it i.e.}, by defining a radial distance $r_j$ from the centre and a polar angle $\theta_j$, where the integer $1 \leq j \leq N$ labels the different spheres. Each sphere possesses a dipole moment $\mathbf{m}_j$ of magnitude $m$ radially tilted away from the normal by an angle $\alpha_j$. In other words, the dipole moment projected onto the horizontal surface lies in the radial direction, either inwards or outwards. The confining field is generated by $N^\prime$ equally spaced magnets of moment $\mathbf{m}^{\mathrm{ext}}$ with magnitude $m^{\mathrm{ext}}$ (see the red arrows in Fig. \ref{f:set-up}(b)), attached to the wall of the container at a radial distance $R$ and polar angles $\theta_k = 2\pi k/N^\prime$ with $k \in \{1, \ldots, N^\prime\}$. Since the container has a conical shape, the magnetic moments along the circumference of the bucket have themselves a radial tilting angle $\Delta$. Finally, the asymmetric structure of the floating object seen in Fig. \ref{f:set-up}(a) adds stability against tilting too easily. For a total mass $M$ and a distance $h$ between the centre of mass and centre of buoyancy, the torque required to flip the floating object out of equilibrium is given by $Mgh$, where $g$ is the gravitational acceleration.

Since all magnetic forces present in the experiment originate from the magnetic moments attached either to the floating sphere or to the perimeter of the containing vessel, a mathematical model describing the experimental setup must include the magnetic dipole interaction as the most fundamental ingredient. The interaction energy of two floating spheres with magnetic dipole moments $\mathbf{m}_i$ and $\mathbf{m}_j$, separated by a vector $\mathbf{r}_{ij}=\mathbf{r}_i-\mathbf{r}_j$ is given by~\cite{jackson}
\begin{equation}
    U^\mathrm{int}_{ij} = \frac{\mu_0}{4\pi}\left( \frac{\mathbf{m}_i\cdot\mathbf{m}_j} {r_{ij}^3} - \frac{3(\mathbf{m}_i\cdot\mathbf{r}_{ij})(\mathbf{m}_j\cdot\mathbf{r}_{ij})}{ r_{ij}^5}\right)\,,
    \label{eq:u-dipole}
\end{equation}
where $\mu_0$ is the magnetic permeability of the surrounding medium.  Accordingly, the interaction energy between a floating sphere and a fixed magnet $k$ is given by
\begin{equation}
    U^\mathrm{ext}_{ik} = \frac{\mu_0}{4\pi}\left( \frac{\mathbf{m}_i\cdot\mathbf{m}_k^\mathrm{ext}} {r_{ik}^3} - \frac{3(\mathbf{m}_i\cdot\mathbf{r}_{ik})(\mathbf{m}_k^\mathrm{ext}\cdot\mathbf{r}_{ik})}{ r_{ik}^5}\right),
\end{equation} 
such that the total magnetic energy perceived by each floating magnet is
\begin{equation}
    U^{\mathrm{mag}}_i = \frac{1}{2}\sum_{\substack{j = 1 \\ i\neq j}}^{N}U^{\mathrm{int}}_{ij} + \sum_{k=1}^{N^\prime}U^{\mathrm{ext}}_{ik}\,.
\end{equation}

Finally, the mechanical energy associated with the torque generated by the gravity-buoyancy pair of forces has the form $U^{\mathrm{mec}}_i= Mgh(1 - \cos{\alpha_i})$. Therefore, the total energy of the system is written as
\begin{equation}
    U^{\mathrm{tot}} = \sum_{i=1}^{N} U^\mathrm{mag}_{i} + U^{\mathrm{mec}}_i\,\,. \label{eq:total_energy}
\end{equation}

It is convenient to adopt a dimensionless radial distance $\rho = r/R$ and a dimensionless energy $U = 4\pi R^3 U^{\mathrm{tot}}/(\mu_0 \,m\,m^{\mathrm{ext}})$. In this form, the total energy (Eq. \ref{eq:total_energy}) is then written as 
\begin{widetext}
\begin{align}
    U &= \frac{\lambda}{2}\sum_{\substack{i, j = 1 \\ i\neq j}}^{N} \left[\frac{\cos{\alpha_i}\cos{\alpha_j} -2\sin\alpha_i\sin\alpha_j\cos{\theta_{ij}}}{(\rho_i^2 + \rho_j^2 - 2\rho_i\rho_j\cos{\theta_{ij}})^{3/2}} + \frac{3\rho_i\rho_j\sin\alpha_i\sin\alpha_j\sin^2{\theta_{ij}}}{(\rho_i^2 + \rho_j^2 - 2\rho_i\rho_j\cos{\theta_{ij}})^{5/2}}\right]\nonumber \\
    & + \sum_{i=1}^{N} \sum_{k=1}^{N^\prime} \left[\frac{\cos\Delta\cos{\alpha_i} - 2\sin\Delta\sin\alpha_i\cos{\Theta_{ik}}}{(1 + \rho_i^2 - 2\rho_i\cos{\Theta_{ik}})^{3/2}} + \frac{3\rho_i\sin\Delta\sin\alpha_i\sin^2{\Theta_{ik}}}{(1 + \rho_i^2 - 2\rho_i\cos{\Theta_{ik}})^{5/2}}\right] + \tau\sum_{i=1}^{N}(1 - \cos{\alpha_i})\,, \label{eq:tot_energy_gsa}
\end{align}    
\end{widetext}
where $\theta_{ij} = \theta_i - \theta_j$ and $\Theta_{ik} = \theta_i - \theta_k$. We also have defined the dimensionless parameters $\lambda = m/m^{\mathrm{ext}}$ and $\tau = 4\pi R^3 Mgh/(\mu_0 m\,m^{\mathrm{ext}})$ as the ratios between the confining and interaction energies, as well as between the mechanical and interaction energies, respectively.

Finding the equilibrium values for the floating magnet positions $(\rho_1, \ldots, \rho_{N}, \theta_1, \ldots, \theta_{N})$ and angles $(\alpha_1, \ldots, \alpha_{N})$ involves minimising the total energy in Eq.  \eqref{eq:tot_energy_gsa} for $3N$ degrees of freedom. The phase space becomes too large to be fully probed even for moderately low values of $N$. 
Therefore, we use the generalised simulated annealing (GSA)~\cite{gsa_tsallis_1996, gsa_tsallis_1996b, gsa_eff_2000, gsa_R_2013} method (see Appendix \ref{app-gsa} for details), which enables us to explore this high-dimensional space efficiently. The values of $\lambda$ and $\tau$ in Eq. \eqref{eq:tot_energy_gsa} are only adjustable parameters here and by using the GSA results with $N=2$ we can reproduce the experimental observations very well with $\lambda = 1.2$ and $\tau = 6.0\times 10^2$. The value of $\lambda$ is consistent with our expectations based on the dimensions of the floaters and the confining magnets. Moreover, experiments in which the magnetic tilt is negligible correspond to $\tau \to \infty$. We also define in our simulations a cut-off distance $d/R = 0.01$ to prevent collapse when two floating magnets are very close to one another or a floater approaches the bucket wall (in the experiment spheres touch at $d/R = 0.26$).  

Fig. \ref{fig:mag_cluster_delta_17} shows the GSA-obtained results for magnetic clusters containing $N$ floating objects, from $N = 2$ to $17$. To avoid too congested a figure, we superimpose the calculated and the experimental structures for one case only ($N=5$), which displays an excellent agreement. The ordered structure reported by Mayer and others~\cite{Mayer1878b, RiverosEtal2004, NemoianuEtal2022} is evident in all equilibrium configurations shown in Fig. \ref{fig:mag_cluster_delta_17}, including the formation of concentric rings whose radii increase with $N$. In addition, the angles of tilt $\alpha_j$ of each floating object now appear as an extra feature that can also be obtained from the simulations. These are represented in the figure by radial arrows starting from the centre of the object with a length that is proportional to $\sin \alpha_j$. The structures seen in Fig. \ref{fig:mag_cluster_delta_17} correspond to the lowest possible energies of Eq. \eqref{eq:tot_energy_gsa}, giving the impression that the GSA-minimization procedure is unsuitable to find more than one ordered structure for a given number $N$. However, when monitoring the energy evolution of the GSA search, we identify many such alternative structures only a few steps before the final convergence is reached, suggesting them as meta-stable configurations (see Appendix \ref{app-gsa} for details).

Instead of focusing on individual floating objects, it is convenient to assign them a radius $\rho$ and a tilt angle $\alpha$ associated with the ring they belong to.  
Both calculated and experimental values of $\rho$ and $\alpha$ for different rings are presented in Fig. \ref{fig:exp_validation}. Calculations, shown as solid lines, together with the experimental measurements depicted by symbols, are plotted as a function of the number of floating objects $N$. Whether the structure has a single ring  ($1\leq N \leq 5$),  two separate rings ($6\leq N \leq 16$), or a total of three rings ($N \geq 16$), the agreement between experiment and calculations is remarkable for the radii $\rho$ all the way to $N=17$, even though the values of the parameters $\lambda$ and $\tau$ were obtained using only $N=2$ data (see panel a). However, the same agreement is not seen in the case of $\alpha$ shown in Fig. \ref{fig:exp_validation}(b). The reason for that slightly higher discrepancy will be explained later. In any case, the inset of Fig. \ref{fig:exp_validation}(b) contains the GSA results displaying how the tilt angle $\alpha$ of the outermost ring increases with the  radius $\rho$, thus serving as a suitable function to describe the magnetic texture of the cluster.   

It is important to assess whether the stability of the GSA results is affected by the cutoff distance used in the calculations. This can be verified by introducing a quantity that establishes whether or not the system is in stable equilibrium. Bearing in mind the nature of the dipole interaction in Eq. \eqref{eq:u-dipole}, if objects are too close together, the magnetic energy becomes too large (in magnitude) to be compensated by the mechanical energy. In this case, the floating magnets will flip and coalesce into one amorphous compact cluster. The following quantity $u$ can then be used as a flag to test whether or not the GSA-found solutions are indeed stable:   
\begin{equation}
    u = \frac{1}{\tau}\min\left\{\frac{4\pi R^3 U_i^{\mathrm{mag}}}{\mu_0 \,m\,m^{\mathrm{ext}}}: i \in \{1, \ldots, N\}\right\}\,,
    \label{eq:u-flag}
\end{equation}
such that the equilibrium configuration is stable when $1 > u > 0$. 

\begin{figure*}[!htbp]
    \centering
    \includegraphics[width=\textwidth]{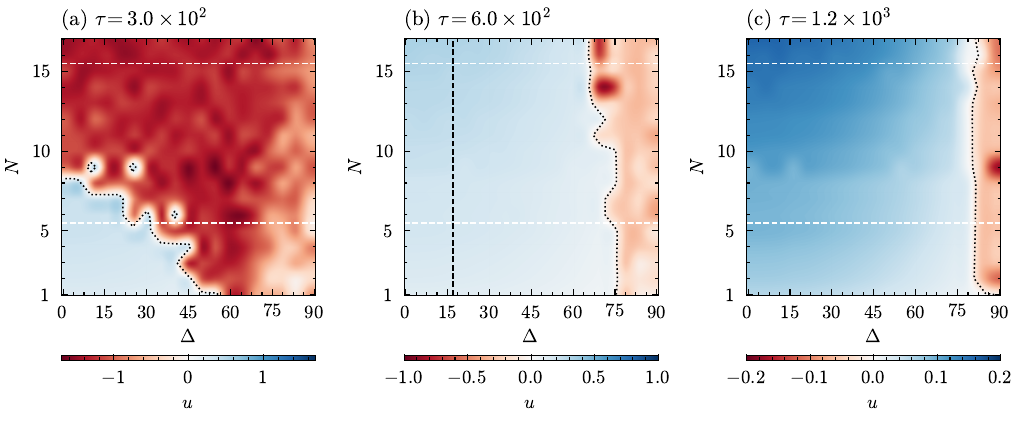}
    \caption{Phase diagrams indicating both the stability and the number of rings in a cluster, depending on the number of floating objects $N$ and the outer angle $\Delta$ (in degrees). The horizontal dashed lines separate the regions with different numbers of rings: the bottom part ($N < 6$) corresponds to cases where only a single ring is formed; the middle ($6\leq N <16$) is for two concentric rings and the top ($N \geq 16 $) is for three. The colour plot depicts the phase space's blue (red) regions where the cluster is mechanically stable (unstable). Black dotted lines mark where in the phase space the system collapses. The three panels are for systems with different degrees of mechanical stability: the middle panel is for $\tau = 6.0 \times 10^2$ obtained from the experiment, whereas the left and right panels are for $\tau = 3.0 \times 10^2$ and $\tau = 1.2 \times 10^3$, respectively. The vertical dashed line in  panel (b) indicates where the experimental measurements were carried out, {\it i.e.}, $\Delta=17^\circ$. }
    \label{fig:phase_diag}
\end{figure*}

The model can now be extended to account for cases not included in the experiment. For example, by varying the confining field direction $\Delta$, by increasing $N$ or by making the objects more or less mechanically stable through a change in $\tau$, we can predict what would occur to the ordered clusters should we use different experimental conditions. This is captured by Fig. \ref{fig:phase_diag} which displays a phase diagram that indicates how many rings are spontaneously self-assembled within a magnetic cluster composed of $N$ floaters exposed to a confining field tilted at an angle $\Delta$. The horizontal dashed lines separate the regions with one, two and three rings. The colour code used in the plot corresponds to the quantity $u$ used to test the stability of the solutions, such that blue (red) indicates the region within the phase space where solutions are stable (unstable) and the three panels are for objects with different values of $\tau$. 

The middle panel of Fig. \ref{fig:phase_diag} is for $\tau=600$, which is the value used to reproduce the experimental results of Figs. \ref{fig:mag_cluster_delta_17} and \ref{fig:exp_validation}. The vertical dashed line seen in that panel at $\Delta =  17^\circ$ indicates that the system is indeed stable and would not collapse for $N$ up to $17$. Moreover, if we were to fix the number of floaters, say at $N=10$, and increase the confining angle continuously, we find that the system would remain with two concentric rings but would no longer be stable when $\Delta$ exceeds $70^\circ$. By carrying out a similar analysis in the case of a cluster composed of mechanically more stable objects ($\tau=1.2 \times 10^3$) we see in Fig. \ref{fig:phase_diag}(c) that the system would remain stable for confining angles of up to approximately $80^\circ$, for all values of $N \leq 20$. In contrast, for the case of smaller values of $\tau$ shown in panel (a), there is a much-reduced region of the phase space for which stable solutions can be found. This is mainly limited to single-ring clusters with confining angles $\Delta < 45^\circ$. A second ring would be possible but for much shallower angles ($\Delta < 20^\circ$) with $N$ not exceeding 8 floaters. 

\section{Effective Single-Ring Model}
\label{esrm}

The phase diagrams of Fig. \ref{fig:phase_diag} are essential to map the possible solutions that magnetic clusters may have in these conditions but it is still computationally onerous to generate them since it involves minimising $3N$ degrees of freedom. Furthermore, because of the form of Eq. \eqref{eq:tot_energy_gsa}, it is difficult to see the relationship between all the relevant quantities in a mathematically transparent way. This calls for a simplified approach, capable of estimating some of these quantities in a less computationally demanding manner. For that purpose, we have derived an effective single-ring model (see Appendix \ref{app-c} for a complete derivation) which provides a set of analytical expressions for the radial distance $\rho$ and tilt angle $\alpha$ of a given ring. They are 
\begin{align}
    \rho &= \left(\dfrac{\lambda}{4\sqrt{2} \cos{\Delta}}\frac{\sum_{i=1}^{N}\sum_{j\neq i}^{N} (1 - \cos{\Tilde{\theta}_{ij}})^{-3/2}}{\sum_{i=1}^{N}\sum_{k=1}^{N^\prime} (5\cos^2{\Tilde{\Theta}_{ik}} - 1)}\right)^{1/5}\,, \label{eq:rho_star} \\
    \alpha &= \frac{3\rho\sin{\Delta}}{\tau}\frac{1}{N}\sum_{i=1}^{N}\sum_{k=1}^{N^\prime}(3\cos^2{\Tilde{\Theta}_{ik}} - 1)\,, \label{eq:alpha_star}
\end{align}
where $\Tilde{\theta}_{ij} = 2\pi(i-j)/N$ and $\Tilde{\Theta}_{ik} = 2\pi(i/N - k/N^\prime)$. 
In the spirit of an effective theory, $\lambda$ and $\tau$ can be assigned as effective parameters selected to match the observed radius and angles to clusters with more than one ring.

\begin{figure}[!htbp]
    \centering
    \includegraphics[width=\columnwidth]{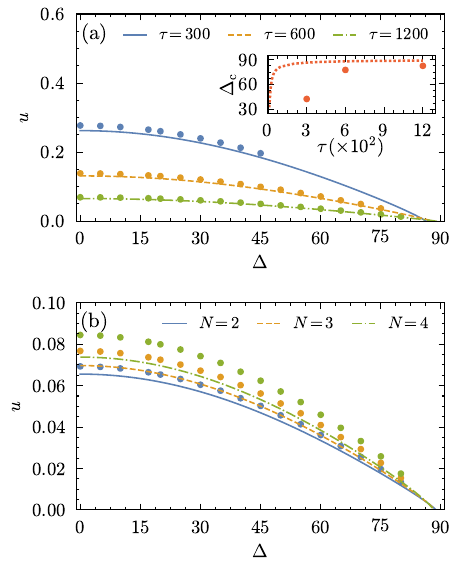}
    \caption{Comparison between the GSA calculations (dots) and the effective single-ring model results (lines) for the quantity $u$, which vanishes when the system is on the verge of collapsing. (a) $u$ is plotted as a function of $\Delta$ (in degrees) for a fixed number of floating objects ($N=2$) and for the three values of $\tau$ used to generate Fig. {\ref{fig:phase_diag}}. The inset plots the critical angle $\Delta_{\mathrm{c}}$ for which $u=0$ but now as a function of continuous values of $\tau$; (b) $u$ as a function of $\Delta$ (in degrees) for $\tau=1.2 \times 10^3$ and for different values of $N$.}
    \label{fig:critical_delta}
\end{figure}

The first indication that Eqs. \eqref{eq:rho_star} and \eqref{eq:alpha_star} are indeed good approximations can be seen in the plots of $\rho(N)$ and $\alpha(N)$ shown as black dotted lines in Figs. \ref{fig:exp_validation}(a) and \ref{fig:exp_validation}(b). They represent the radii and tilt angles of the outermost ring for the cases $N=1$ to $N=17$ and agree well with the GSA results. Furthermore, the dependence of $\rho$ and $\alpha$ on the confining angle $\Delta$, which is difficult to extract from Eq. \eqref{eq:tot_energy_gsa}, becomes a lot more transparent, {\it i.e.}, $\rho \propto (1/\cos\Delta)^{1/5}$ and $\alpha \propto (\tan \Delta)^{1/5} \times (\sin \Delta)^{4/5}$. This is further evidenced in Appendix \ref{app-c}. Another important point that can be seen more transparently in Eq. \eqref{eq:alpha_star} is the inter-dependence of these two quantities such that $\alpha$ depends on $\rho$ and vice-versa, corroborating our earlier  point that the inclusion of the magnetic tilt angle as an extra degree of freedom does impact the equilibrium structure of the cluster.  

This effective single-ring model can also be used to predict the transition to collapse as shown in Fig. \ref{fig:critical_delta}. Panel (a) plots the energy $u$, defined in Eq. \eqref{eq:u-flag}, as a function of $\Delta$ for the same $\tau$ values used to generate Fig. \ref{fig:phase_diag} but in this case the number of floating objects is fixed at $N=2$. The dots are the GSA results and correspond to the $\Delta$ values used to generate all three phase diagrams whereas the continuous lines are the $u$ values obtained from Eqs. \eqref{eq:rho_star} and \eqref{eq:alpha_star}. With such a good agreement, we can ask when the system will collapse by imposing the critical condition $u=0$. Rather than being limited to the computationally time-consuming results of Fig. \ref{fig:phase_diag}, this enables us to establish the critical angle $\Delta_{\mathrm{c}}$ above which the system collapses for a continuous range of $\tau$ values, seen in the inset. The agreement between the points and the dotted line in the inset improves as $\tau$ increases, which is compatible with the assumptions made in the derivation of our effective model. Note that $\Delta_c(\tau \to \infty) = 90^\circ$. Alternatively, we can keep $\tau$ fixed and vary $N$ instead. In this case, Fig. \ref{fig:critical_delta}(b) shows that for $\tau=1.2 \times 10^3$ all $u(\Delta)$ curves cross the $u=0$ line roughly at the same value, regardless of how many floaters $N$ the system has. Once again, this is in agreement with what has been previously seen in Fig. \ref{fig:phase_diag}(c).

Finally, the effective single-ring model can assist us in explaining why the agreement between the GSA calculations and the measured values of $\alpha$ seen in Fig. \ref{fig:exp_validation}(b) are not as striking as for the radii shown in Fig. \ref{fig:exp_validation}(a). The second order derivatives of the energy $U$ with respect to the radii and to the tilt angle  reflect how much energy is required to take the system out of their equilibrium positions in $\rho$ and $\alpha$, respectively. Once again, by making use of Eqs. \eqref{eq:rho_star} and \eqref{eq:alpha_star} the total energy $U$ can be obtained in a much simpler form, which in turn enables us to obtain the derivatives in question in terms of their associated Hessian matrix $\mathbf{H}$ (see Appendix \ref{app-c} for details). We find that the second-order derivative of $U$ with respect to $\rho$ is considerably larger than that associated with $\alpha$, which means that the system is a lot more resistant to changes in $\rho$. In other words, the floating objects resist more strongly to variations in their radial positions but are floppier with regard to changes in the tilt angles $\alpha$. That is one source of explanation for the discrepancy seen in Fig. \ref{fig:exp_validation}(b). Another possible source for the discrepancy is in the sensibility of the GSA results to the introduction of symmetry-breaking fluctuations. For example, by assuming non-uniform $\Delta$ values for the confining magnets, the perfect radial symmetry is broken, giving rise to fluctuations both in $\rho$ and $\alpha$. As seen in Appendix \ref{app-b}, the standard deviation in the tilt angle $\alpha$ is approximately 100 times larger than that found for $\rho$, which adds weight to the argument that one may require a far more precise measurement apparatus to capture the distribution for the tilt angle $\alpha$. 

\section{Conclusions and outlook}

Having modified and adapted Mayer's iconic experiment~\cite{Mayer1878a,Mayer1878b}, we have described how mutually repulsive particles in the presence of a confining field are spontaneously ordered, but this time we have fully accounted for the magnetic tilt that 
arises as these dipole-moment-carrying particles interact. In particular, we focus on how this additional degree of freedom may impact the cluster structure as a whole. We control the naturally occurring tilt between interacting dipole moments by making the cluster objects more mechanically stable. We find that the angle $\Delta$ of the confining magnetic field plays a key role in defining the cluster radius, the magnetic texture described by $\alpha(\rho)$ and the overall stability conditions of dipole clusters. A mathematical model based on the energy balance resulting from the competition between the magnetic forces and the counteracting mechanical torque provides excellent agreement with observations. 

More than just reproducing the experiment, we believe the model contains the essential ingredients to describe the properties of both ordered and amorphous clusters with minimal modifications. First and foremost, the GSA technique seems suitable to handle much larger clusters. Changes to the confining geometry~\cite{cluster_geometries} can be easily accounted for without increases to the computational costs. Because the model is not exclusive of the macroscopic system considered here but valid across a wide range of length scales, it may be adapted and used to describe the case of dipole-coupled nanoparticles. By breaking the radial symmetry considered here, a much richer magnetic texture is likely to appear which, in turn, will affect how a cluster responds to external fields. Finally, we have not considered the dynamic aspects of the cluster formation ~\cite{Krishnamurthy_2022} in the presence of tilted magnetic fields but the current work provides all the necessary ingredients to address that.



\section*{Acknowledgements}
The authors thank Karl Morgan for helpful discussions. 
P. D. S. de Lima acknowledges the CAPES/PRINT program (process no. 88887.838338/2023-00). J. M. de Araújo acknowledges the CNPq Brazilian research agency for funding (grant 311589/2021-9).

\appendix
\counterwithin{figure}{section}

\section{List of stable and metastable configurations}
\label{app-structures}

Table \ref{tab:structures} lists all the structures (up to N=17) that have been reported in the comprehensive studies by both Mayer~\cite{Mayer1878b} and Nemoianu {\em et al.}~\cite{NemoianuEtal2022}, together with our own experimental and computational findings, and some numerical results in Ref. \cite{RiverosEtal2004}. The notation $(n_1, n_2, n_3, ...)$ refers to the number of magnets $n_\ell$ in ring $\ell$, with $\ell=1$ being the innermost ring (entries $n_\ell=0$ are not displayed). Our GSA simulations were primarily aimed at identifying structures of minimal energy, however, we also recorded some meta-stable structures, corresponding to local energy minima, see for example Fig. \ref{fig:gsa} for the case $N=5$. They are consistent with those seen in previous experiments and simulations, as indicated in Tab. \ref{tab:structures}. 

Curiously, for particle number $N$ up to and including $N=14$, the minimum energy configurations that we identified for our floating magnets are identical to those found in simulations of two very different systems. These are, firstly, two-dimensional Coulomb clusters, interacting via a repelling Yukawa type potential and confined in a parabolic potential \cite{LaiLin1999}, and secondly, equal-area partitions of a circle with minimal total line length~\cite{CoxFlikkema2010} (i.e. a monodisperse foam confined by a circular perimeter) .


\begin{table*}[!htbp]
    \centering
    \begin{ruledtabular}
    \begin{tabular}{l l l}
        $N$ & Stable & Metastable\\
        \hline
        1 & $(1)^{s,e}$~\cite{Mayer1878b,NemoianuEtal2022} & - \\
        2 & $(2)^{s,e}$~\cite{Mayer1878b,NemoianuEtal2022} & - \\
        3 & $(3)^{s,e}$~\cite{Mayer1878b,NemoianuEtal2022} & - \\
        4 & $(4)^{s,e}$~\cite{Mayer1878b,NemoianuEtal2022} & - \\
        5 & $(5)^{s,e}$~\cite{Mayer1878b,NemoianuEtal2022} & $(1,4)^{s,e}$~\cite{Mayer1878b}\\
        6 & $(1,5)^{s,e}$ \cite{Mayer1878b,NemoianuEtal2022,RiverosEtal2004} & $(6)^{e}$ \cite{Mayer1878b,NemoianuEtal2022,RiverosEtal2004}\\
        7 & $(1,6)^{s,e}$ \cite{Mayer1878b,NemoianuEtal2022,RiverosEtal2004} & - \\
        8 & $(1,7)^{s,e}$~\cite{Mayer1878b} &  $(2,6)$ \cite{Mayer1878b} \\
        9 & $(2,7)^{s,e}$~\cite{Mayer1878b,NemoianuEtal2022} & $(1,8)^{e}$ \\
        10 & $(2,8)^{s,e}$ & $(3,7)^{s}$~\cite{Mayer1878b,NemoianuEtal2022} \\
        11 & $(3,8)^{s,e}$~\cite{Mayer1878b,NemoianuEtal2022} & $(2,9)^{e}$\\
        12 & $(3,9)^{s,e}$~\cite{RiverosEtal2004,NemoianuEtal2022} & $(4,8)^{s,e}$~\cite{Mayer1878b,NemoianuEtal2022}\\
        13 & $(4,9)^{s,e}$~\cite{Mayer1878b,NemoianuEtal2022} & $(3,10)^{ e}$, $(5,8)$~\cite{NemoianuEtal2022} \\
        14 & $(4,10)^{s,e}$ & $(5,9)^{s,e}$~\cite{Mayer1878b,NemoianuEtal2022}, $(1,5,8)$ ~\cite{NemoianuEtal2022}, $(1,6,7)$~\cite{NemoianuEtal2022}\\
        15 & $(4,11)^{s,e}$ & $(5,10)^{s, e}$~\cite{RiverosEtal2004}, $(1,5,9)^{s}$~\cite{Mayer1878b}, $(1,6,8)$~\cite{NemoianuEtal2022} \\
        16 & $(1,5,10)^{s,e}$~\cite{RiverosEtal2004} & $(5,11)^{e}$, $(1,7,8)$~\cite{NemoianuEtal2022}, (1,6,9) \cite{Mayer1878b} \\
        17 & $(1,5,11)^{s,e}$ & $(5,12)^{e}$, $(1,6,10)^{e}$~\cite{Mayer1878b, NemoianuEtal2022}\\
    \end{tabular}
    \end{ruledtabular}
    \caption{List of structures found for up to $N=17$ magnets in confinement. The superindices $s$ and $e$ refer, respectively, to the simulations and experiments described in the main text.}
    \label{tab:structures}
\end{table*}

\section{Generalized Simulated Annealing}
\label{app-gsa}
\begin{figure}[!htbp]
    \centering
    \includegraphics[width=\columnwidth]{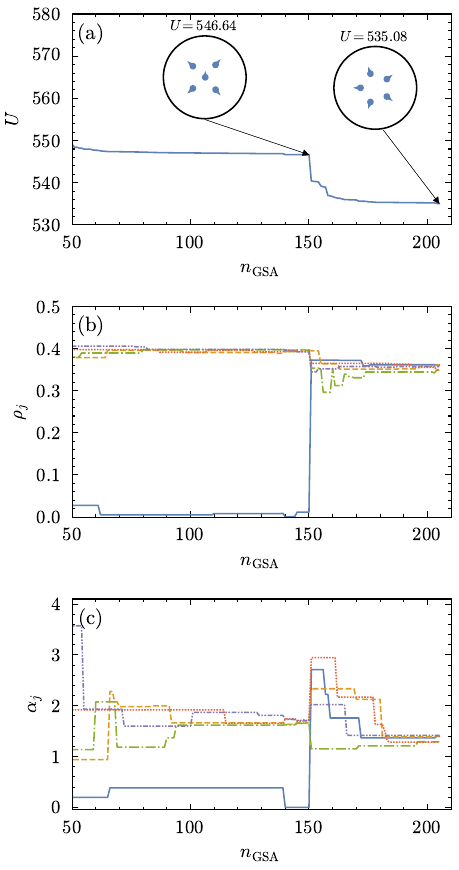}
    \caption{Illustration of the GSA history for $N=5$ using the calibrated values of $\lambda = 1.2$ and $\tau = 6.0\times 10^2$ for $\Delta = 17^\circ$. Panel (a) shows the minimization of the energy defined in Eq. \eqref{eq:tot_energy_gsa} after 50 GSA steps where two possible ordered clusters are displayed as insets. Panels (b) and (c) present the radii $\rho_j$ and tilting angles $\alpha_j$ (in degrees) evolution for each one of the five floating objects.}
    \label{fig:gsa}
\end{figure}

We use generalized simulated annealing (GSA) to minimize the total energy Eq. \eqref{eq:tot_energy_gsa} by parameterising it in terms of a $3N$-dimensional vector $\boldsymbol{\Xi}$ with coordinates $\Xi_j = (\rho_j, \theta_j, \alpha_j)$. The GSA then performs $t \in \{1, \ldots, n_{\mathrm{GSA}}\}$ steps starting from an initial random state $\boldsymbol{\Xi}_{t = 0}$ at temperature $T(t = 0)$ which is moved to a new state $\boldsymbol{\Xi}_{t = 1} = \boldsymbol{\Xi}_{t = 0} + \Delta \boldsymbol{\Xi}$ where the perturbation $\Delta\boldsymbol{\Xi}$ obeys the visiting distribution~\cite{gsa_tsallis_1996}
\begin{align}
            g_{q_v}&(\Delta\boldsymbol{\Xi}) = \left(\frac{q_v - 1}{\pi}\right)^{d/2}\frac{\Gamma\left(\frac{1}{q_v-1} + \frac{d-1}{2}\right)}{\Gamma\left(\frac{1}{q_v-1} - \frac{1}{2}\right)}T_{q_v}^{d/(q_v-3)}\nonumber\\
            &\,\times \left[1 + (q_v - 1)     \frac{(\Delta\boldsymbol{\Xi})^2}{T_{q_v}^{2/(3-q_v)}}\right]^{1/(1-q_v) + (1-d)/2}\,,
\end{align}
which is a distorted Cauchy-Lorentz distribution of dimension $d = 3N$, written in terms of Tsallis' numbers $q_v$ and $q_a$. These numbers define the visiting $T_{q_v}$ and acceptance $T_{q_a}$ temperatures. We use a \texttt{SciPy} implementation~\cite{scipy} with $q_v = 2.62$ and $q_a = -5.00$~\cite{gsa_eff_2000}. The new state $\boldsymbol{\Xi}_{t = 1}$ is accepted with probability
\begin{equation}
            p_a = \min{\left(1, \left[1 + \frac{(q_a - 1)\Delta U_t}{T_{q_a}}\right]^{1/(1-q_a)}\right)}\,,
\end{equation}
where $\Delta U_t$ is the energy difference between these two states. Once accepted, the temperatures are cooled following the scheme
\begin{equation}
    T_{q_v}(t) = \frac{T_{q_v}(0)(2^{q_v-1} - 1)}{(1 + t)^{q_v - 1} - 1}\,, \quad T_{q_a}(t) = T_{q_a}(0)/t\,,
\end{equation}
where $T_{q_v}(0)$ and $T_{q_a}(0)$ are the initial temperatures. In this regard, the GSA behaves like the classical simulated annealing for high temperatures and as the steepest descent algorithm for low temperatures.

For all simulations presented in this manuscript, we run the GSA by $n_{\mathrm{runs}} = 10$ times from different initial random states using up to $n_{\mathrm{GSA}} = 1.0\times 10^{19}$ steps.  After that, the result with the lowest positive $u$ calculated using Eq. \eqref{eq:u-flag} was selected. We note that when $u < 0$ for a given run, the other instances also indicate collapse. For example, Fig. \ref{fig:gsa} depicted the steps in which the energy decreased for $N = 5$ using the experiment-calibrated values of $\lambda$ and $\tau$. The energy evolution in Fig. \ref{fig:gsa}(a) shows a possible local minimum after $\sim 150$ GSA steps, where we see an arrangement with one floating magnet at the centre and the remaining magnets ordered on a second ring. However, thanks to the local search behaviour of GSA at lower temperatures, a ground energy state can be found after $\sim 200$ steps with the five floaters now on a unique ring. This cluster geometry change is also seen in Figs. \ref{fig:gsa}(b) and Fig. \ref{fig:gsa}(c), where we note an abrupt variation of the radial distance $\rho_j$ and tilting angle $\alpha_j$ for one of the five floaters before reaching the most energetic favourable equilibrium configuration. Both ordered structures appearing in the insets of Fig. \ref{fig:gsa}(a) are indeed observed in the experiment, supporting the idea of two local minima. This apparent multiplicity of equilibrium structures is also experimentally found for other cases (see Tab. \ref{tab:structures}). Therefore, the GSA is not only useful to provide the true ground state configuration but is also capable of pointing to the existence of meta-stable states that may be close in the energy space. 

\section{Cluster robustness against symmetry-breaking fluctuations}
\label{app-b}

We consider a more general scenario where the confining field generated by the $N^\prime$ fixed magnets around the perimeter has random fluctuations in their tilt angles $\Delta_k$. In that case, the problem is no longer radially symmetric and the polar coordinate of $\mathbf{m}_j$ associated with the horizontal surface projection does not necessarily coincide with the floater's polar coordinate $\theta_j$. Therefore, we need to assign an additional angle,  which we label $\beta_j$, to account for that. The total energy of Eq. \eqref{eq:total_energy} is now written in its non-radial form as the $4N$-dimensional function
\begin{widetext}
\begin{align}
    &U^{\mathrm{non-rad}} = \frac{\lambda}{2}\sum_{\substack{i, j = 1 \\ i\neq j}}^{N} \left[\frac{\cos\alpha_i\cos\alpha_j + \sin\alpha_i\sin\alpha_j\cos{\beta_{ij}}}{(\rho_i^2 + \rho_j^2 - 2\rho_i\rho_j\cos{\theta_{ij}})^{3/2}} - \frac{3\sin\alpha_i\sin\alpha_j(\rho_i\cos{\omega_{ii}} - \rho_j\cos{\omega_{ji}})(\rho_i\cos{\omega_{ij}} - \rho_j\cos{\omega_{jj}})}{(\rho_i^2 + \rho_j^2 - 2\rho_i\rho_j\cos{\theta_{ij}})^{5/2}}\right]\nonumber \\
    + &\sum_{i=1}^{N} \sum_{k=1}^{N^\prime} \left[\frac{\cos{\Delta_k}\cos{\alpha_i} + \sin{\Delta_k}\sin\alpha_i\cos{\omega_{ki}}}{(1 + \rho_i^2 - 2\rho_i\cos{\Theta_{ik}})^{3/2}} - \frac{3\sin\Delta_k\sin\alpha_i(\rho_i\cos{\omega_{ii}} - \cos{\omega_{ki}})(\rho_i\cos{\omega_{ik}} - 1)}{(1 + \rho_i^2 - 2\rho_i\cos{\Theta_{ik}})^{5/2}}\right] + \tau\sum_{i=1}^{N}(1 - \cos{\alpha_i})\,, \label{eq:tot_energy_non-rad}
\end{align}    
\end{widetext}
where $\omega_{ij} = \theta_i - \beta_j$. We use the same GSA recipe described in Appendix \ref{app-gsa} but now for an augmented parameter space given by $\Xi_j = (\rho_j, \theta_j, \alpha_j, \beta_j)$ coordinates.

\begin{figure}[!htbp]
    \centering
    \includegraphics[width=\columnwidth]{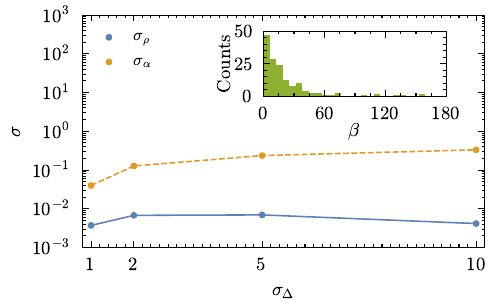}
    \caption{Standard deviation of the radii $\rho$ (blue dot) and of the tilt angle $\alpha$ (orange dot) as a function of the standard deviation of the outer angle $\Delta$. This corresponds to calculated values for the outermost ring and $N=6$, for which the mean value of $\Delta$ is kept at $17^\circ$. The inset shows the $\beta$ angle distribution observed in the experiment. This angle measures the deviation from the radial direction.}
    \label{fig:fluctuations}
\end{figure}

Fluctuations in the $\Delta_k$ angles are related to inherent random errors of experimental realization of attaching fixed magnets on the bucket surface. To test that, we consider four situations where these fluctuations are normally distributed around $\Delta = 17^\circ$ having distinct dispersion levels $\sigma_\Delta$. Fig. (\ref{fig:fluctuations}) shows the impact of these fluctuations on the standard deviations of the radius and of the tilt angles, both for the case of the outermost ring of $N=6$. The results corroborate the earlier posed claim of higher robustness for the radius $\rho$ against symmetry-breaking fluctuations in $\Delta$, compared to the same effect on the tilt angles. One more evidence of such symmetry-breaking fluctuations can be seen by the observed angles that were not  tilted exactly along the radial direction. This appears in the distribution of the angle $\beta$ that measures the deviations from the radial direction. Although the distribution displayed in the inset of Fig.\ref{fig:fluctuations} is peaked at $\beta=0$ thus  indicating a predominant radial tilt angle, it does have a non-zero width that confirms the presence of deviations that may be as large as $60^\circ$.

\section{Single-ring model derivation}
\label{app-c}

\begin{figure}[!htbp]
    \centering
    \includegraphics[width=\columnwidth]{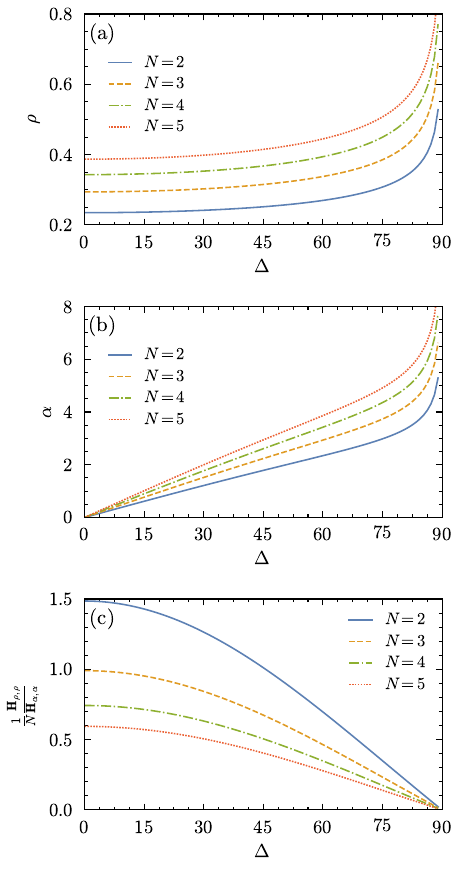}
    \caption{Equilibrium radius $\rho$ and tilting angle $\alpha$ obtained using the single-ring approximation model are shown in panels (a) and (b). These values are calculated as a function of $\Delta$ for different $N$ values. Panel (c) shows the curvature ratio for the same cases.}
    \label{fig:analytical_solution}
\end{figure}

We derive a semi-analytical solution for the equilibrium configuration of floating magnets having the same radius $\rho$ and tilting angle $\alpha$. In this regime, the total energy Eq. \eqref{eq:tot_energy_gsa} can be rewritten as a two-dimensional function
\begin{widetext}
\begin{align}
    U^{\mathrm{sym}} = &\sum_{i=1}^{N}\sum_{k=1}^{N^\prime} 
    \left(\frac{\cos{\Delta}\cos{\alpha} - 2\cos{\Tilde{\Theta}_{ik}}\sin{\Delta}\sin{\alpha}}{(1 + \rho^2 - 2\rho\cos{\Tilde{\Theta}_{ik}})^{5/2}}
    + \frac{3\rho\sin^2{\Tilde{\Theta}_{ik}}\sin{\Delta}\sin{\alpha}}{(1 + \rho^2 - 2\rho\cos{\Tilde{\Theta}_{ik}})^{3/2}}\right) \nonumber \\ 
    + \frac{\lambda}{8\sqrt{2}}&\frac{1}{\rho^3}\sum_{\substack{i, j = 1 \\ i\neq j}}^N \left(\frac{2}{(1 - \cos{\Tilde{\theta}_{ij}})^{3/2}} +\frac{\sin^2{\alpha}}{(1 - \cos{\Tilde{\theta}}_{ij})^{1/2}}\right) 
    + N\tau(1 - \cos{\alpha})\,, \label{eq:tot_energy_sym}
\end{align}
\end{widetext}
where $\Tilde{\theta}_{ij} = 2\pi(i/N-j/N)$ and $\Tilde{\Theta}_{ik} = 2\pi(i/N - k/N^\prime)$. For $\lambda \sim 1$ the competition between confining and floating interaction potentials makes the objects concentrate around the centre of the bucket, meaning that $\rho$ is small. The maximum radial distance experimentally verified is $\rho \sim 0.6$. Moreover, $\alpha \leq 15^\circ$ in all cases for $\tau \sim 10^2$, indicating a small-angle regime. Such observations allow us to employ a parabolic approximation for the confining potential and a quadratic expansion in the tilting angle-dependent expressions of Eq. \eqref{eq:tot_energy_sym}. Under these considerations, the symmetrized total energy Eq. \eqref{eq:tot_energy_sym} becomes
\begin{align}
    U^{\mathrm{sym}} &\approx  
    A(\alpha^2 + 2)\rho^2 +  B\rho^2\alpha + C(\alpha^2 - 2)\rho + D(\alpha^2 - 2) \nonumber\\ + &E\rho\alpha + F\alpha
    + \frac{1}{\rho^3}\left(G + H\alpha^2\right) + \frac{N\tau}{2}\alpha^2\,, \label{eq:tot_energy_quadratic}
\end{align}
where the confining potential coefficients are given by
\begin{align}
    A &= \frac{3}{4}\cos{\Delta}\sum_{i=1}^{N}\sum_{k=1}^{N^\prime}(5\cos^2{\Tilde{\Theta}_{ik}} - 1)\,, \label{eq:a}\\
    B &= 6\sin{\Delta}\sum_{i=1}^{N}\sum_{k=1}^{N^\prime}(3\cos{\Tilde{\Theta}_{ik}} + 5\cos^3{\Tilde{\Theta}_{ik}})\,, \\
    C &= -\frac{3}{2}\cos{\Delta}\sum_{i=1}^{N}\sum_{k=1}^{N^\prime}\cos{\Tilde{\Theta}_{ik}}\,, \\
    D &= -\frac{1}{2}NN^\prime\cos{\Delta}\,, \\
    E &= 3\sin{\Delta}\sum_{i=1}^{N}\sum_{k=1}^{N^\prime}(1 - 3\cos^2{\Tilde{\Theta}_{ik}})\,, \label{eq:f}\\
    F &= -2\sin{\Delta}\sum_{i=1}^{N}\sum_{k=1}^{N^\prime}\cos{\Tilde{\Theta}_{ik}}\,,
\end{align}
while the interaction potential is parameterized by
\begin{align}
    G &= \frac{\lambda}{4\sqrt{2}}\sum_{\substack{i, j = 1 \\ i\neq j}}^N(1 - \cos{\Tilde{\theta}_{ij}})^{-3/2}\,,\label{eq:u}\\
    H &= \frac{\lambda}{8\sqrt{2}}\sum_{\substack{i, j = 1 \\ i\neq j}}^N(1 - \cos{\Tilde{\theta}_{ij}})^{-1/2}\,.
\end{align}

Therefore, the equilibrium points are obtained by taking the partial derivatives of Eq. \eqref{eq:tot_energy_sym}  with respect to $\rho$ and $\alpha$, giving rise to the pair of coupled nonlinear equations
\begin{align}
    (2A + C - 3H/\rho^4)\alpha^2 &+ (2B\rho + E)\alpha \nonumber\\
    &= 3G/\rho^4 - 2(2A - C)\,, \label{eq:partial_rho}\\    
    (2A\alpha + B)\rho^2 &+ (2C\alpha + E)\rho \nonumber \\
    &= -F - (D + N\tau)\alpha\,, \label{eq:partial_alpha}
\end{align}
where a linear relation between $\rho$ and $\alpha$ can be obtained from Eq. \eqref{eq:partial_alpha} 
\begin{equation}
    \alpha \approx -\rho E/N\tau\,, \label{eq:alpha_rho}
\end{equation}
also providing an inverse dependence between $\alpha$ and $\tau$. Moreover, this approximation holds since $E>B>F$ and $\rho < 1$. Accordingly, a quintic equation for the equilibrium $\rho$ can now be acquired by replacing Eq. \eqref{eq:alpha_rho} into Eq. \eqref{eq:partial_rho}, which dropping $1/\tau$ and $1/\tau^2$ terms, turns to be 
\begin{equation}
    4A\rho^5 - 2C\rho^4 = 3G\,,
\end{equation}
which for $2A \gg -C$ has the following solution
\begin{equation}
    \rho = \left(3G/4A\right)^{1/5}\,,\label{eq:rho_analytical_appendix}
\end{equation}
where the equilibrium $\alpha$ is calculated using the result given by Eq. \eqref{eq:rho_analytical_appendix} in Eq. \eqref{eq:alpha_rho}. Panels (a) and (b) of Fig. \ref{fig:analytical_solution} show how this equilibrium solution varies with $\Delta$ for small values of $N$, indicating slower variation of $\rho$ when compared to $\alpha$. Moreover, the expressions appearing in the main text, Eqs. \eqref{eq:rho_star} and \eqref{eq:alpha_star}, are obtained by replacing the definitions given in Eqs. \eqref{eq:a} and \eqref{eq:f} for $A$ and $E$ and in Eq. \eqref{eq:u} for $G$. Similarly, the energy value Eq. \eqref{eq:tot_energy_sym} for this equilibrium configuration can be evaluated. To conclude, we argue that this single-ring approximation is also useful to explicit the local curvature around the equilibrium positions Eqs. \eqref{eq:alpha_rho} and \eqref{eq:rho_analytical_appendix} by keeping the assumptions of small $\rho$ and $\alpha$. In that case, the (Hessian) dynamical matrix is given by~\cite{stability_2, stability_1}
\begin{equation}
    \mathbf{H}_{\rho, \alpha} = 
    \begin{pmatrix}
        4A + 12G/\rho^5 & E \\
        E & 2D + N\tau
    \end{pmatrix}\,,
\end{equation}
with diagonal elements
\begin{widetext}
\begin{align}
    \mathbf{H}_{\rho, \rho} &= 3\cos{\Delta}\sum_{i=1}^{N}\sum_{k=1}^{N^\prime}(5\cos^2{\Tilde{\Theta}_{ik}} - 1) + \frac{3\lambda}{\sqrt{2}}\frac{1}{\rho^5}\sum_{\substack{i, j = 1 \\ i\neq j}}^N(1 - \cos{\Tilde{\theta}_{ij}})^{-3/2}\,, \label{eq:curv_rho}\\
    \mathbf{H}_{\alpha, \alpha} &= N(\tau - N^\prime\cos{\Delta})\,. \label{eq:curv_alpha} 
\end{align}
\end{widetext}

It is evident from Eqs. \eqref{eq:curv_rho} and \eqref{eq:curv_alpha} the role played by $\Delta$ in the cluster stability. Floaters closer to the centre are more radially stable due to $\sim 1/\rho^5$ dependence in Eq. \eqref{eq:curv_rho}, while $\tau$ is fundamental to guarantee the tilt angle stability as seen in Eq. \eqref{eq:curv_alpha}. Moreover, Eq. \eqref{eq:curv_alpha} provides a lower bound value for $\tau$. To see that, consider the limit $\Delta \rightarrow 0$ where the Hessian matrix is truly diagonal. In this case, $\tau > N^\prime$ to produce stable equilibrium regardless the value of $N$. The ratio between these curvatures as a function of $\Delta$ is shown in Fig. \ref{fig:analytical_solution}(c), where stability for the radial distance remains higher than the tilting angle up to $\Delta\sim 70^\circ$.

\bibliography{references-magnets}
\end{document}